\documentclass[twocolumn,showpacs,preprintnumbers,amsmath,amssymb]{revtex4}
\usepackage{graphicx}% Include figure files
\usepackage{dcolumn}% Align table columns on decimal point
\usepackage{bm}% bold math

\begin{document}

%\preprint{APS/123-QED}

\title{Expansion of a dipolar condensate}
\author{S. Yi and L. You}

\affiliation{School of Physics, Georgia Institute of Technology,
Atlanta, GA 30332-0430, USA}

\begin{abstract}
We discuss the expansion of an initially trapped dipolar
condensate. When the nominal isotropic s-wave interaction strength
becomes tunable through a Feshbach resonance,
anisotropic dipolar effects are shown to be
detectable under current experimental conditions [E. A. Donley
{\it et al.}, Nature {\bf 412}, 295 (2001)].
\end{abstract}

\date{\today}

\maketitle

Atomic Bose-Einstein condensates (BEC) are realized at extremely
low temperatures \cite{bec,mit,rice}, when short range atom-atom
interaction can be described by a single parameter: the s-wave
scattering length $a_{\rm sc}$. This is a valid approximation as
all higher order partial wave collisions die away in a
short range potential when the collision energy approaches zero.
In more realistic models, however, even
ground state atoms may be polarized such as in an external
magnetic trap, when the direction of its spin (of the valence
electron for alkali) becomes aligned. BEC with dipole-dipole
interaction has attracted considerable attention in recent
years \cite{yi1, goral1, santos, yi2, goral2, pavel}, and
trapped fermionic dipoles have also been studied \cite{goral3}.

Despite the fact that dipolar interactions are now widely
known to exist in a trapped BEC, it has
not been directly detected yet. This is because
the dipole interaction is much weaker than the contact
interaction under most circumstances.
Recently we studied the small amplitude
shape oscillation of a dipolar condensate \cite{yi2}.
When the s-wave scattering length
$a_{\rm sc}$ is tuned near to zero as realizable in
the $^{85}$Rb experiment \cite{cornish,donley},
we showed that the anisotropic dipolar effect becomes
detectable under current experimental conditions.

The free expansion of an interacting condensate after its
sudden release from the trap is now a standard
diagnostic tool in BEC physics \cite{castin,holland}.
It has also been extended to the case of an interacting Fermi gas
\cite{menotti}. In this paper, we investigate the
free expansion of a dipolar condensate \cite{goral3}.
Our results show that this may also lead to
the experimental detection of dipolar interactions.
This paper is organized as follows; First we briefly
review the formulation for a dipolar BEC; we then
discuss our study of its expansion
with both variational and numerical methods.
We will subsequently concentrate on a detailed
study for situations corresponding to the $^{85}$Rb
experiment \cite{donley}. Then we
conclude with some new results on the
shape and stability of a dipolar condensate.

For simplicity, we study a trapped dipolar BEC assuming
all atomic dipole moments are equal and aligned along
the z-axis.
The atom-atom interaction is then simplified to be
\begin{eqnarray}
V(\vec R)=g_0\delta(\vec R)+g_2{1-3\cos^2\theta_R\over R^3}
\end{eqnarray}
where $\vec R=\vec r-\vec r\,'$, $\theta_R$ is the polar angle of
$\vec R$, $g_0=4\pi\hbar^2a_{\rm sc}/M$
representing the contact interaction, and in atomic units
$g_2=\alpha^2(0){\cal E}^2$ ($\alpha(0)$ atomic polarizability) or
$\mu^2$ ($\mu$ magnetic dipole moment). The corresponding
Gross-Pitaevskii equation (in adimensional form) is then
\begin{eqnarray}
i\dot{\psi}(\vec r)&=&\left[-{\nabla^2\over 2}+V_{\rm
ext}(\vec r) +\sqrt{(2\pi)^3}P_0|\psi(\vec
r)|^2\right.\nonumber\\&+&\left.{3\over 2}\sqrt{2\pi}P_2\int d\vec
r\,' {1-3\cos^2\theta\over R^3} |\psi(\vec
r\,')|^2\right]\psi(\vec r),\hskip 24pt
\label{gpe}
\end{eqnarray}
where $\psi(\vec r)$ is the condensate wave function
(normalized to $1$) and a cylindrical symmetric harmonic trap
in dimensionless unit $V_{\rm ext}(\vec
r)=(x^2+y^2+\lambda^2z^2)/2$, with an aspect ratio $\lambda$.
The length
unit is $a_\perp=\sqrt{\hbar/M\omega_\perp}$
and energies are measured in units of
$\hbar\omega_\perp$ ($\omega_\perp$ is the
radial frequency of the trap). $P_0=\sqrt{2\pi}Na_{\rm
sc}/a_\perp$ measures the contact interaction
strength while
$P_2=\sqrt{2\pi}Ng_2/(3\hbar\omega_\perp a_\perp^3)$
denotes the strength of the
dipole-dipole interaction. The ground state
wave function of a dipolar condensate can be found by replacing
the lhs of Eq. (\ref{gpe}) with $-\mu\psi$, where $\mu$ is the
chemical potential.

To proceed with the study of the condensate
free expansion; we can solve the Eq. (\ref{gpe}) numerically.
We initialize the wave function to the self-consistently solved
ground state in the presence of the trap $V_{\rm ext}$,
then find the time evolved wavefunction from the Eq. (\ref{gpe})
by employing a self-adaptive Runge-Kutta method without
the $V_{\rm ext}$. In practice, this becomes an expansive
calculation as we we have to use a rather large spatial
grid to accommodate the ever-expanding wave function and
to obtain sufficient accuracy. We find that it is necessary
to check the accuracy of solutions repeatedly on different sized
grids.

Alternatively, the variational approach as developed
earlier \cite{yi2} can be used. In this case, we assume that
the wave function always takes the form of a Gaussian,
and transform the Equation (\ref{gpe}) into the
following equations for variational parameters
\begin{eqnarray}
\ddot{q}_r+q_r&=&{1\over q_r^3}-{1\over q_r^3q_z}[P_2f(\kappa)-P_0],\\
\ddot{q}_z+\lambda^2q_z&=&{1\over q^3_z}-{1\over
q_r^2q^2_z}[P_2g(\kappa)-P_0],
\label{eqndyna}
\end{eqnarray}
which in fact represent the
radial and axial condensate widths of the expanding condensate
as $q_r=\sqrt{\langle x^2\rangle/2}=\sqrt{\langle
y^2\rangle/2}$ and $q_z=\sqrt{\langle z^2\rangle/2}$.
$\kappa=q_r/q_z$ is the condensate aspect ratio,
generally different from the trap aspect ratio $\lambda$,
and $f(\kappa)=
[-4\kappa^4-7\kappa^2+2+9\kappa^4H(\kappa)]/2(\kappa^2-1)^2$,
$g(\kappa)=[-2\kappa^4+10\kappa^2+1-9\kappa^2H(\kappa)]/(\kappa^2-1)^2$
with $H(\kappa)={\tanh}^{-1}\sqrt{1-\kappa^2}/\sqrt{1-\kappa^2}$.
The kinetic and interaction energy per atom can then
be expressed in terms of condensate widths
according to
\begin{eqnarray}
E_{\rm kin} &=& E_{\rm kin}^{(r)}+E_{\rm kin}^{(z)}\nonumber\\
&=&{1\over 2}({1\over q_r^2}+\dot{q_r}^2)
+{1\over 4}({1\over q_z^2}+\dot{q_z}^2),\nonumber\\
E_{\rm int} &=& E_{\rm sc}+E_{\rm dd}\nonumber\\
&=&{P_0\over 2q_r^2q_z}+P_2{2\kappa^2+1-3\kappa^2H(\kappa)
\over 2q_r^2q_z(\kappa^2-1)} .
\end{eqnarray}

\begin{figure}
\centering
\includegraphics[width=3.25in]{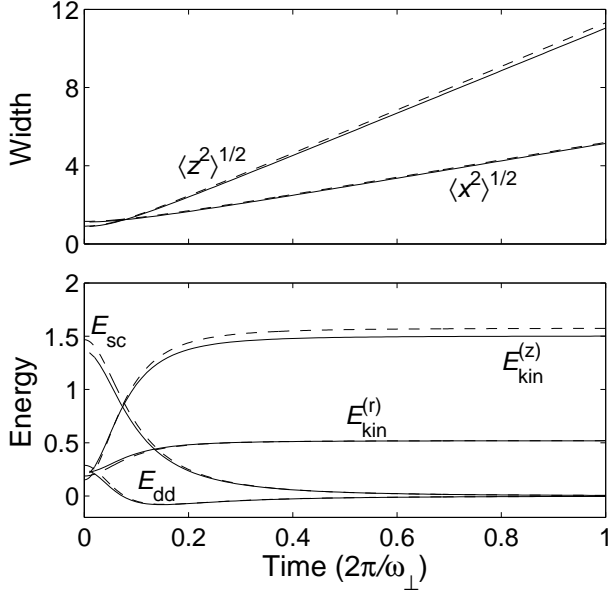}
\caption{Comparisons of free expansion results
from the numerical (solid line) and the variational
(dashed line) approaches as discussed in the text,
for $\lambda=2$, $P_0=10$, and $P_2=10$.
The upper panel is for the condensate
widths (in units of $a_\perp$) and the lower panel for the
various energy components (in units of $\hbar\omega_\perp$).}
\label{figfreevarnumdipl2p10}
\end{figure}

The validity of the variational solution has been checked
thoroughly for the frequencies
of condensate small amplitude shape oscillations \cite{santos,yi2}.
We have checked over a wide range of parameters before
concluding that it is also justified for use in the
free expansion problem. This enabled us tremendous freedom
in exploring the expansion dynamics without resorting
to the time consuming numerical solutions.
Figure \ref{figfreevarnumdipl2p10}
shows the time dependence of condensate widths and energy
components assuming $\lambda=2$ and $P_0=P_2=10$. We see that the
variational approach indeed gives a very good approximation to the
numerical calculations. This result is also confirmed for
$\lambda$ from $0.1$ to $3$.

\begin{figure}
\centering
\includegraphics[width=3.25in]{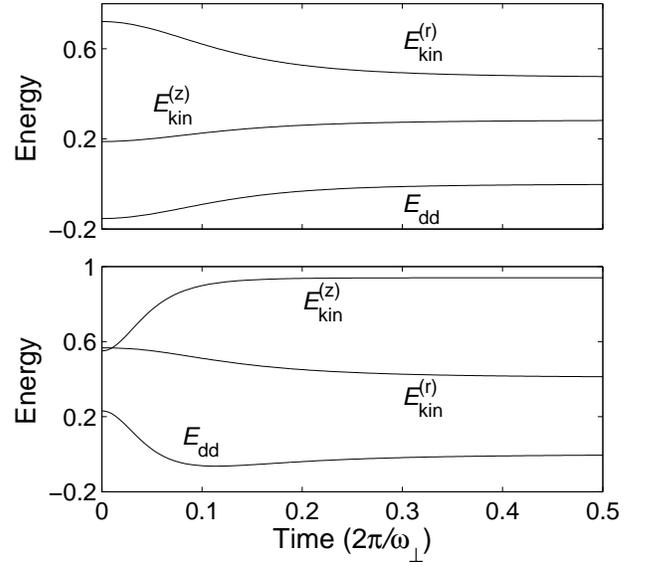}
\caption{The time dependence of various energy components
(in units of $\hbar\omega_\perp$) during free expansion
for $\lambda=1$
(upper panel) and $3$ (lower panel). Other parameters are $P_0=0$
and $P_2=1$.
} \label{figfreeenergy}
\end{figure}

The time dependent behavior of the energy components is
rather interesting. As shown in Fig. \ref{figfreeenergy},
irrespective of the initial signs of the overall
dipole interaction energy, it always approaches
zero asymptotically during the expansion.
In the early stages, however, its
behavior depends on trap geometry
due to the anisotropic nature of the interaction. For small $\lambda$, the
initial dipole-dipole interaction energy is negative; during the
expansion, it always gains energy from the kinetic energy and
approaches zero monotonically. For large $\lambda$ when the
initial dipole-dipole
interaction energy is positive, the condensate first releases
its interaction energy until $E_{\rm dd}$
becomes negative, then gains energy and finally
approaches zero.

The effects of dipolar interaction on the kinetic energies
are rather simple. We find that, independent of $\lambda$,
the dipolar interaction always decreases $E_{\rm kin}^{(r)}$
while increases $E_{\rm kin}^{(z)}$ with time,
i.e., the dipole-dipole interaction causes
the transfer of radial kinetic energy into the axial direction.
This phenomenon is observed even when $P_0\neq 0$.
This result contradicts intuition,
because along the z-axis, the dipole-dipole
interaction is attractive, one would expect that the
$E_{\rm kin}^{(z)}$ would decrease with time because the atoms
are being slowed down due to the dipolar attraction.
To resolve this puzzle, we note that the kinetic energy
along radial or axial directions each has two parts:
one from the gradient of the wave function ($1/2q_r^2$)
which decreases with time, and the other from the expanding
gas ($\dot{q}^2/2$) which increases with time. This observed
phenomena shows that in the radial direction the increase of
the kinetic energy due to $\dot{q}_r^2/2$ cannot
compensate for the decrease due to $1/2q_r^2$.
Similarly, we can understand the net increase of kinetic
energy along the axial direction.

For the remainder of this paper, we will focus our studies on
$^{85}$Rb condensate as in the JILA experiment
\cite{cornish,donley}, where the tuning of the scattering length has
been demonstrated in a remarkable fashion utilizing the Feshbach resonance.
In this case, the magnetic dipole moment of the trapped state
$|F=2,M_F=2\rangle$ is $\mu=2\mu_B/3$ ($\mu_B$ is the Bohr magneton).
We adopt the radial frequency
$\omega_\perp=2\pi\times 17.35$ (Hz) as from the experiment
\cite{cornish,donley} and assume that the
asymmetric parameter $\lambda$ can be adjusted.
The resulting dipolar interaction strength
is $P_2=5.0\times 10^{-6}N$.

The release energy $E_{\rm rel}=E_{\rm kin}+E_{\rm int}$ is
the total energy of the condensate after switching off the
trapping potential. It's values can be strongly affected by
the atom-atom interaction.
For non-interacting atoms, the release energy per atom
$E_{\rm rel}^0=(1+\lambda/2)/2$, is independent of the
atom number. For interacting atoms, it depends on the
total number of atoms since both
$P_0$ and $P_2$ are proportional to $N$. If the interaction is
repulsive, the release energy per atom always increases with $N$.
For attractive interaction, the release energy always decreases
(increases) with $N$ if $N$ is above (below) some critical value.
Depending on the geometry of the
trapping potential, the overall dipole-dipole interaction
can be either repulsive or attractive, and the release energy
also shows different behaviors for different trapping potentials.

\begin{figure}
\centering
\includegraphics[width=3.25in]{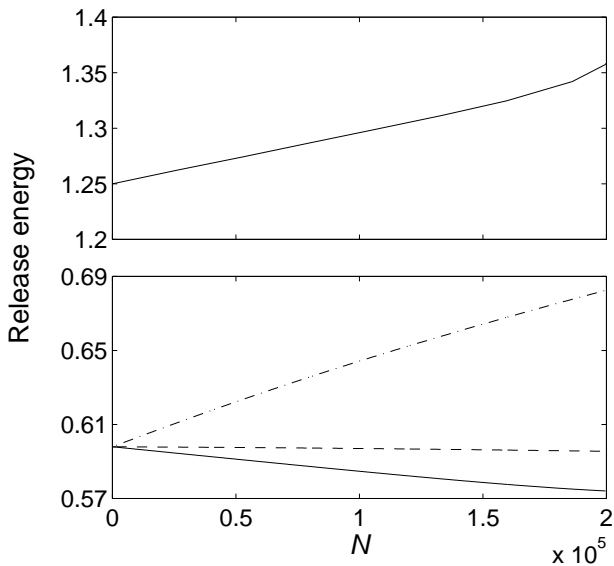}
\caption{The atom number dependence of the total release energy
for a $^{85}$Rb condensate with $\lambda=3$ (upper panel) and
$\lambda=6.8/17.35$ (lower panel), for the s-wave scattering length
$a_{\rm sc}=0$ (solid line), $0.1$ (dashed line), and $0.5$ (dash-dotted
line) ($a_B$).} \label{figrelpc}
\end{figure}

In the upper panel of Fig. \ref{figrelpc},
we present the atom number dependence of
the release energy for $\lambda=3$. Since the dipole-dipole
interaction is predominately repulsive in this case,
the release energy increases with $N$.
In the lower panel for $\lambda=6.8/17.35$,
when $P_0=0$ (solid line), the release energy decreases with $N$.
We find that this behavior holds even for a very small
positive scattering length as shown in Fig. \ref{figrelpc}.
As one might have expected, only for very small values
of the s-wave scattering length, are the effects of
dipole-dipole interaction important.

\begin{figure}
\centering
\includegraphics[width=3.25in]{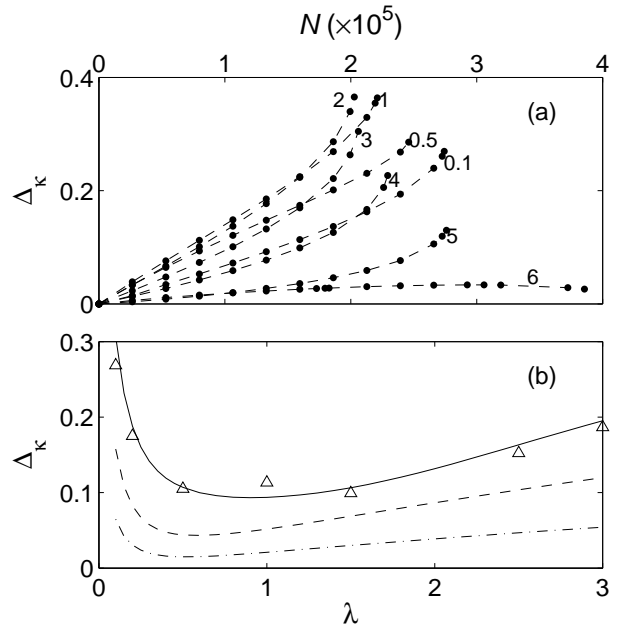}
\caption{The relative change of condensate aspect ratio due to
the dipole interaction. (a) For the trapped ground state
with $\lambda=0.1$, $0.5$, $1$, $2$, $3$, $4$, $5$, and $6$.
(b) The asymptotic condensate aspect ratio
for $N=2\times 10^5$ (solid line), $10^5$ (dashed line), and
$4\times 10^4$ (dash-dotted line). Triangle markers show
the results for $N=2\times 10^5$ from numerically calculated
expansions.} \label{figrelkappau}
\end{figure}

The dipolar interaction can also change the condensate aspect ratio.
To quantify this effect, we use the relative difference of
the (ground state) condensate aspect ratio
\begin{eqnarray}
\Delta_\kappa ={|\kappa(P_2\neq 0
)-\kappa(P_2=0)|\over \kappa(P_2=0)}.
\end{eqnarray}
In Fig. \ref{figrelkappau} (a),
we plot the numerical results of $\Delta_\kappa(P_0=0,P_2)$ for
various values of $\lambda$.
We see that the relative differences can
become as high as 30\% for a wide range of
trap aspect ratio $\lambda$.
An experimental measurement of these differences will
represent a direct detection of dipolar effects,
although the following technical difficulty remains;
The condensates produced in
current $^{85}$Rb experiments \cite{donley}
contain rather small number of atoms that a
{\it in-situ} direct optical imaging is very challenging.
The free expansion as discussed in this paper,
leads to larger condensate sizes as shown
in Fig. \ref{figfreevarnumdipl2p10}, thus allowing
for easier imaging of condensate aspect ratios.
We find that the condensate aspect ratio changes during
the expansion and eventually approaches a constant value,
which we call the asymptotical aspect ratio.
Not surprisingly, this asymptotical aspect ratio
also depends on the dipole-dipole interaction.
In Fig. \ref{figrelkappau} (b), we present the $\lambda$ dependence
of the relative change of the asymptotical condensate ratios for
various values of $N$. We see that for carefully chosen parameters
$\Delta_\kappa(P_0=0,P_2)$ can become large enough to be observed
experimentally.

\begin{figure}
\centering
\includegraphics[width=3.25in]{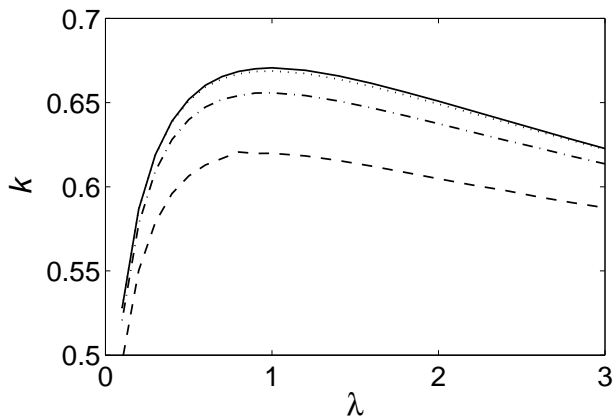}
\caption{The effects of dipole-dipole interaction on the
stability coefficient $k$ for a $^{85}$Rb condensate
with scattering length
$a_{\rm sc}=-0.5 a_0$ (dashed line), $-1.0a_0$ (dash-dotted line),
and $-3.0a_0$ (dotted line). The solid line indicates the
stability coefficient when dipole-dipole interaction is not
included.} \label{figcolpcneg}
\end{figure}

Finally, as already studied extensively before,
the partially attractive nature of the dipole-dipole interaction
can destabilize the condensate ground state in traps with small values of
$\lambda$ ($<5.2$) \cite{yi1,santos}. The stability coefficient
$k=N_{cr}|a_{\rm sc}|/a_{\rm ho}$
(with $a_{\rm ho}=\sqrt{\hbar/m\bar\omega}$ and
$\bar\omega=(\omega_x\omega_y\omega_z)^{1/3}$)
is frequently used to measure the stability of a condensate \cite{gammal}.
In Fig.
\ref{figcolpcneg}, we plot the $\lambda$
dependence of $k$ for various scattering lengths
as obtained from variational study. We see that for a small negative
scattering length, the dipole-dipole interaction changes the
$k$ value dramatically, another signature for directly detecting
the dipolar interaction.

In conclusion, we have studied the free expansion of a
dipolar condensate. We have shown that the weak dipole-dipole
interaction may become detectable through several
observations, of which the measurement of the condensate
aspect ratio after an expansion period seems the most promising.
We also discussed briefly the effects of the dipolar interaction
on destabilizing the condensate when the nominal s-wave
scattering is tuned close to zero, a situation close to the
recent experimental measurement of $k$ in $^{85}$Rb condensates \cite{roberts}.

We thank Dr. Raman for insightful discussions.
This work is supported by the NSF grant No. PHYS-0140073.

\end{document}